\documentclass[11pt,twoside]{article}

\usepackage{asp2006}
\usepackage{epsf}
\usepackage{lscape}

\def\lesssim{\mathrel{\hbox{\rlap{\hbox{\lower4pt\hbox{$\sim$}}}\hbox{$<$}}}}
\def\gtrsim{\mathrel{\hbox{\rlap{\hbox{\lower4pt\hbox{$\sim$}}}\hbox{$>$}}}}

\begin{document}
\title{Turmoil in Orion: The Nearest Massive Protostar}   
\author{Jonathan C. Tan}   
\affil{Dept. of Astronomy, University of Florida, Gainesville, FL 32611, USA}    

\begin{abstract} 
I discuss different theories of massive star formation: formation from
massive cores, competitive Bondi-Hoyle accretion, and protostellar
collisions. I summarize basic features of the Turbulent Core Model
(TCM). I then introduce the Orion Kleinmann-Low (KL) region, embedded
in the Orion Nebula Cluster (ONC) and one of the nearest regions of
massive star formation. The KL region contains three principal radio
sources, known as ``{\it I}'', ``{\it n}'' and ``BN''. BN is known to
be a runaway star, almost certainly set in motion by dynamical
ejection within the ONC from a multiple system of massive stars, that
would leave behind a recoiling, hard, massive, probably eccentric
binary. I review the debate about whether this binary is $\Theta^1C$,
the most massive star in the ONC, or source {\it I}, and argue that it is
most likely to be $\Theta^1C$, since this is now known be a recoiling,
hard, massive, eccentric binary, with properties that satisfy the
energy and momentum constraints implied by BN's motion. Source {\it n} is a
relatively low-mass protostar with extended radio emission suggestive
of a bipolar outflow. Source {\it I}, located near the center of the main
gas concentration in the region, the Orion Hot Core, is the likely
location of a massive protostar that is powering the KL region, and I
discuss how its basic properties are consistent with predictions from
the TCM. In this scenario, the radio emission from source {\it I} is the
base of a bipolar outflow that is ionized by the massive protostar and
should be elongated along the axis of the outflow.
\end{abstract}

\section{Introduction and Definitions}

Understanding the formation of massive stars is an important problem
for many areas of astrophysics including high redshift Population III
star formation, galaxy formation and evolution, galactic center
environments and supermassive black hole formation, star and star
cluster formation, and planet formation around stars in clusters,
which may be relevant to our own solar system. The problem is
challenging because of the wide range of spatial and temporal scales,
the complicated interplay of gravity, thermal pressure, magnetic
fields, radiation and ``turbulent'' (i.e. nonthermal) motions,
including bipolar outflows from surrounding stars, and the uncertain
initial conditions, caused by high obscuration to the typically
distant and crowded regions where massive stars form. Thus progress
requires close testing of theoretical models against detailed
observational data, such as is available for the nearest massive
star-forming regions.

I adopt the following definitions (see also McKee \& Tan 2003; Beuther
et al. 2007): a {\it protostar} is a star (i.e. a near-equilibrium
gaseous configuration in which gravity is balanced by thermal and
radiation pressure and possibly rotation) accreting matter so that its
mass, $m_*$, has not yet reached its maximum value, $m_{*f}$, that it
is born with; a {\it massive protostar} (MP) or star has a mass $\geq
8 M_\odot$; a {\it pre-(massive-)stellar core} is the self-gravitating
(negative total energy), topologically-connected, matter surrounding
the location where the (eventually massive) protostar first forms (at
time $t=0$); at later times when the protostar and
rotationally-supported disk have formed, this system is a {\it
(massive-)star-forming core}, which includes the mass of the protostar
and disk (mass may join or leave the core during the growth of the
protostar); in principle the evolution of the pre-stellar core at
earlier times ($t<0$) could be followed, defining its center to be at
the minimum of the gravitational potential; a {\it star-forming clump}
or {\it protocluster} is the system made up of a forming cluster of
stars and the self-gravitating gas surrounding them. The minimum
number of stars, which may be protostars, to define a cluster, ${\cal
N}_{\rm *,min}$, can be debated: I suggest ${\cal N}_{\rm *,min} = 10
\gg 1$. For clumps, self-gravity should first be assessed for all the
mass in the minimum convex volume that includes the ${\cal N}_*$
stars, then surrounding stars and gas (including effects of external
pressure) can be assessed to see if they are bound to, and therefore a
part of, this structure.

Note, since the mass of a protostar when it first forms is expected to
be very small ($\ll M_\odot$), massive protostars must have gone
through a stage when they had low and intermediate masses. Beuther et
al. (2007) refer to these as ``Low to Intermediate Mass Protostars
destined to become Massive Protostars'' and here we suggest the
abbreviation LIMP-MP, so the expected evolutionary sequence if massive
star formation is a scaled-up version of low-mass star formation is
Pre-Massive-Stellar Core $\rightarrow$ Massive-Star-Forming Core
containing LIMP-MP $\rightarrow$ Massive-Star-Forming Core containing
MP $\rightarrow$ Massive Star. Much of the theoretical debate in the
field centers on whether the pre-massive-stellar core is itself {\it
massive} (i.e. with mass comparable to or larger than $m_{*f}$),
whether the massive-star-forming core containing the LIMP-MP is {\it
massive}, and whether there is a stage where the mass of the
massive-star-forming core containing the MP is gas dominated. The
alternative is that the core has a small mass when the protostar first
forms and then accumulates the bulk of its mass during the
star-forming stage while maintaining a relatively small gas mass fraction.

Star-forming clumps, which are typically supersonically turbulent,
form new cores and protostars by ``gravitational fragmentation'' (but
note that the material in the new cores will still be part of the
clump, so the clump itself has not strictly speaking fragmented). New
cores and protostars may also form inside existing star-forming
cores. I distinguish ``disk fragmentation'' from ``turbulent
fragmentation''. In disk fragmentation, which only occurs inside
star-forming cores, the new core (and protostar) is in a low
eccentricity orbit about the original protostar, i.e. is
gravitationally bound to it with infall mostly resisted by rotational
support, and is likely to have its boundaries set by tidal forces from
the original protostar. The original star-forming core maintains its
identity, but now with a binary and circumbinary disk at its
center. Turbulent fragmentation (e.g. Padoan \& Nordlund 2002) can
occur both inside and outside of star-forming cores. If it takes place
inside a star-forming core, the new core will be bound to the previous
core (it is part of it), but will not have its infall supported
significantly by rotation. One expects that the enclosed mass for the
position of the new core inside the old core will be dominated by the
protostar in the case of disk fragmentation and (non-stellar) gas in
the case of turbulent fragmentation, but these are not discrete
categories.

If a star-forming core undergoes turbulent fragmentation, it is
possible to view it as forming a sub-cluster within the main
protocluster. This could be a major difference from the isolated mode
of star formation that forms single stars (or binaries or higher
multiples via disk fragmentation). Nevertheless the system can
approximate that of isolated star formation from cores if a
significant fraction, say $\geq 1/2$, of the {\it stellar} mass
produced in the core is in a single star (or binary or higher multiple
formed by disk fragmentation). In this situation there is still,
approximately, a one-to-one correspondence of the core with the final
principal star. Thus, another important discriminator between star
formation models is the relative importance of turbulent fragmentation
that occurs in gas that is already part of a star-forming core
compared to gas that is part of the clump, but not yet associated with
any particular protostar.  It should be noted that since massive stars
are rare (they contain a small fraction of the total young stellar
mass), pre-massive-stellar cores are also rare: they require
relatively special conditions to form.

Pre-stellar cores may form in a clump that is already rich in
(proto)stars so that, especially for large, massive cores, some stars
are embedded in the core volume. The orbits of these stars in the
clump may be such that they are not bound to the new core, so that
their mass is not formally part of the core by the above definitions,
even though their mass played some typically minor role in defining
its potential. As these stars orbit through and out of the core they
may accrete some mass by Bondi-Hoyle accretion, with gas streamlines
typically being intercepted by a pre-existing remnant disk from an
earlier accretion phase. This Bondi-Hoyle accretion of the core gas is
a continuation of Bondi-Hoyle accretion of clump gas that these
protostars would have been experiencing prior to the formation of the
core.  Note that Bondi-Hoyle accretion involves accretion of gas that
is initially not bound to the protostar (strictly speaking not bound
to the star-forming core, which is composed mostly of the protostar,
together with a remnant accretion disk). The mass affecting the
gravitational cross-section here is dominated by the protostar.  This
is to be contrasted with accretion of gas to cores in which the gas
mass fraction of the core is still significant and the gas
distribution still extended. This latter case is less likely to be
affected by protostellar feedback.

What if a pre-existing star is bound to the new core? Most stars in
these environments will still have at least some remnant accretion
disk and so will be protostars and thus would have had remnant cores
of their own. This situation can be viewed in a number of ways. From
the point of view of a new pre-massive-stellar core, it is just about
to form with its new central, defining protostar (that later will
become a massive star), but finds itself infested with older,
pre-existing stars that may steal some of the massive core gas. From
the point of view of the pre-existing protostars, they will likely see
their boundaries grow, and maybe merge, as the massive core forms or
grows around them, but their boundaries would not include the new
LIMP-MP\footnote{A pre-massive-stellar core could
in principle form inside a star-forming core by fragmentation, but
then would be expected to have a low mass and to not typically contain
pre-existing stars.}. 
They would likely experience enhanced accretion, possibly contributing
significantly to their final mass.  

Given these possibilities, one basic question to answer is ``how do
massive protostars typically build up their mass?'' The turbulent core
model (McKee \& Tan 2002, 2003) posits that massive stars form from
massive, gas-dominated, relatively-near-equilibrium cores, that form
by turbulent fragmentation from the magnetized clump medium or by
pre-stellar-core agglomeration processes, have some significant
fraction of their support from turbulent motions and that during their
collapse channel a large fraction of the star-forming gas via a
central disk into just one star (or a few stars formed by disk
fragmentation). 

An alternative possibility is competitive Bondi-Hoyle accretion
without involving massive gas dominated cores around the massive
protostar (Bonnell et al. 2001; Schmeja \& Klessen 2004; Bonnell,
Vine, \& Bate 2004; Bonnell \& Bate 2006). The gas in the core is a
relatively small fraction of the massive protostellar mass, but is
continuously being replenished by efficient accretion of gas that was
previously not bound to the protostar. In these numerical models, this
gas is typically being funneled to the center of a star-forming clump
that is undergoing global collapse. For some of the massive stars
formed in these models -- those whose protostellar seeds are amongst
the first to form in the cluster -- their formation can be described
as involving a massive gas core, but one which subsequently undergoes
efficient turbulent fragmentation to put most of its mass into a
cluster of low-mass stars. In neither of these descriptions is there a
close correspondence between massive core mass and resulting massive
star mass. Since these simulations only include thermal pressure and
no magnetic pressure, it is not surprising that they do not see
massive near-equilibrium cores, with masses much greater than the
thermal Jeans mass, which is $<M_\odot$ in typical massive
star-forming regions.

A third possibility, requiring extremely high stellar densities, is
the growth of massive protostars by stellar mergers (Bonnell et
al. 1998), which is effectively a merger of star-forming cores,
followed by the merger of the protostars. Bally \& Zinnecker (2005)
invoked this mechanism to explain the ``explosive'' nature of the
outflow from the Orion KL region (Allen \& Burton 1993), although we
shall see that another more likely possibility is the tidally-enhanced
accretion and accretion-powered outflow from close passage of a
fast-moving star (BN) with a massive protostar and accretion disk
(source {\it I}) (Tan 2004).

\section{The Turbulent Core Model}

McKee \& Tan (2002; 2003, hereafter MT03) modeled massive star
formation by assuming an initial condition that is a marginally
unstable, massive, turbulent core in approximate pressure equilibrium
with the surrounding protocluster medium, i.e. the star-forming
clump. This clump was also assumed to be in approximate hydrostatic
equilibrium so that its mean internal pressure is $P\sim G\Sigma^2$,
where $\Sigma = M/(\pi R^2)$ is the mean mass surface density with typical
observed values $\sim 1\:{\rm g\:cm^{-2}}$. This pressure sets the
overall density normalization of each core and thus its collapse time
and accretion rate. The core density structure adopted by MT03 is
$\rho \propto r^{-k_\rho}$, with $k_\rho=1.5$ set from
observations. This choice affects the evolution of the accretion rate:
$k_\rho<2$ implies accretion rates accelerate. However, this is a
secondary effect compared to the overall normalization of the
accretion rate that is set by the core's external pressure. 

Since much of the pressure support in the core is nonthermal with
significant contributions from turbulent motions, one does not expect
a smooth density distribution in the collapsing core, and the
accretion rate can show large variations about the mean. Also the
assumption that the core is collapsing in isolation is of course
approximate: MT03 estimate that during the collapse the core interacts
with a mass of surrounding clump gas similar to its initial mass, although not
all of this will become bound to the core. The mass spectrum of cores
may be shaped by core agglomeration and disruption processes. The
former will be more efficient in the dense centers of clumps, perhaps
leading to more frequent massive core and massive star formation in
these regions.


Predictions of the TCM are the properties of the cores and accretion
disks of massive protostars. The initial core size is $R_{\rm core}
\simeq 0.06 (M_{60})^{1/2}\Sigma^{-1/2}\:{\rm pc}$, where
$M_{60}=M_{\rm core}/60\:M_\odot$. Note an allowance has been made for
massive cores tending to be near the centers of clumps, where
pressures are about twice the mean (MT03).  These cores have
relatively small cross-sections for close interactions with other
stars. The accretion rate to the star, via a disk, is $\dot{m}_* =
4.6\times 10^{-4} f_*^{1/2} M_{60}^{3/4}\Sigma^{3/4}\:M_\odot\:{\rm
yr}^{-1}$, where $f_*$ is the ratio of $m_*$ to the final stellar mass
and a 50\% formation efficiency due to protostellar outflows is
assumed, so $m_{*f}=0.5 M_{\rm core}$. The collapse time, $1.3\times 10^5 M_{60}^{1/4}
\Sigma^{-3/4}\:{\rm yr}$, is short and quite insensitive to $m_{*f}$,
allowing coeval stochastic high and low-mass star formation in a
cluster, that might take $\gtrsim 1$~Myr to build up.  The disk size
is $R_{\rm disk}=1200 (\beta/0.02) (f_* M_{60})^{1/2}
\Sigma^{-1/2}{\rm AU}$, where $\beta$ is the initial ratio of
rotational to gravitational energy of the core, and the normalization
is taken from typical low-mass cores (Goodman et al. 1993), although
there is quite a large dispersion about this value.

These estimates of the accretion rate allow quantitative models of the
protostellar evolution, allowing prediction of the stellar radius
$r_*(m_*)$, luminosity $L_*(m_*)$, H-ionizing luminosity $S_*(m_*)$,
disk structure and outflow intensity, which can then be compared to
observed systems (see Figure 1 of Tan 2003). For example, a
$20\:M_\odot$ protostar accreting from an originally $60\:M_\odot$
core near the center of a $\Sigma=1\:{\rm g\:cm^{-2}}$ clump would
have $L_* \simeq 10^5\:L_\odot$ and $S_* \sim 10^{48}\:{\rm photons\:
s^{-1}}$. The protostellar outflow should have injected
$\sim5000\:M_\odot {\rm km\:s^{-1}}$ of momentum into the surrounding
gas, enough to eject a substantial fraction of the original core
material in directions above and below the accretion disk. The outflow
can also confine the ionizing luminosity in equatorial directions,
creating an outflow-confined hypercompact HII region (Tan \& McKee
2003).

\section{The Orion KL Region}

The closest massive protostar is thought to be radio source {\it I}
(Menten \& Reid 1995) in the Orion Kleinmann-Low (KL) region,
$414\pm7$~pc away (Menten et al. 2007). This region is near the center
of the Orion Nebula Cluster (ONC), marked by the Trapezium OB
stars. Also nearby is the Becklin-Neugebauer (BN) object, known to
have a high proper motion, equivalent to about 40~$\rm km\:s^{-1}$ in
the plane of the sky (Plambeck et al. 1995; Tan 2004; G\'omez et
al. 2005), and source {\it n}, a relatively low-luminosity, low-mass
protostar (Gezari, Backman, \& Werner 1998).

\subsection{The BN Object: A Runaway Star Ejected from $\Theta^1C$}

BN's luminosity is $\sim 2500-10^4\:L_\odot$ (Gezari et al. 1998),
corresponding to a zero age main sequence B3-B4
(8-12~$M_\odot$) star. It is highly likely that BN originated in the
ONC. Since the cluster is too young for binary supernova ejections,
the most plausible model for BN's motion is dynamical ejection from an
unstable triple or higher order system. This can often occur when a
hard binary interacts with another star (Hut \& Bahcall
1983). Typically the least massive star is ejected at about the escape
speed from the remaining binary at the orbit of the secondary, which
is often left eccentric.

I have proposed BN was ejected from an interaction with the
$\Theta^1C$ system (Tan 2004) because: (1) $\Theta^1C$ lies along
BN's past trajectory; (2) $\Theta^1C$ has a proper motion direction
opposite to BN's (van Altena et al. 1988); (3) $\Theta^1C$ has a
proper motion amplitude that would predict BN's mass is $6.4\pm3
M_\odot$, in agreement with the estimate from its luminosity;
(4) $\Theta^1C$ has a relatively massive ($\gtrsim 6\:M_\odot$)
secondary companion (Schertl et al. 2003) (now known to be
$15.5\:M_\odot$, Kraus et al. 2007); (5) the orbit of the $\Theta^1C$
secondary is now known to be highly eccentric ($e=0.91$, Kraus et
al. 2007; however, this is disputed by Patience et al. 2008); (6) the
semi-major axis of the $\Theta^1C$ binary (total mass $\simeq
50\:M_\odot$) is about 17AU and the escape speed from this distance is
$70\:{\rm km\:s^{-1}}$, high enough to explain BN's speed (the binding
energy of the binary is now known to be $2.6\times 10^{47}\:{\rm
  ergs}$ compared to BN's kinetic energy of $\simeq 1.6\times
10^{47}\:{\rm ergs}$). To have all of the above occur by chance is
highly improbable. Furthermore, no other revealed, massive ONC stars
have any of the correct proper motion or binary properties.

G\'omez et al. (2005) and Rodriguez et al. (2005) proposed that BN was
ejected from an interaction with source {\it I} and source {\it n}
from a location about 4\arcsec to the NW of source {\it I}'s current
position. This is based on the apparent radio proper motions of {\it
I} and {\it n}. However, both these radio sources are elongated along
directions parallel to the claimed proper motion vectors, increasing
the uncertainties in the derived motions. The dense gas and dust that
now appears in the vicinity of source {\it I} on scales $\gtrsim
100$~AU (Blake et al. 1996; Wright et al. 1996; Beuther et al. 2006)
could not have been retained by the star if it had been subject to
such an ejection event. I expect that source {\it I} has a much
smaller proper motion relative to the ONC than has been claimed, and
that it is forming from the surrounding gas core in which it is now
embedded.


\subsection{Source I: Core, Disk, Protostar and Outflow}


Wright et al. (1992) mapped a core of dense gas in the KL region in
emission at 450~$\rm \mu m$ and 3.5~mm: the center of this source is
often referred to as the Orion Hot Core. This core is centered close
($\lesssim 1\arcsec$) to the position of source {\it I} and has a
scale of about 0.05~pc (25\arcsec) across its short axis in the SE to
NW direction.  The core is elongated along the NE to SW axis, and
indeed it likely part of a larger-scale filamentary feature in these
directions. The scale of the core is similar to that expected for an
initially massive $\sim 60\:M_\odot$ core in near equilibrium with a
$\Sigma=1\:{\rm g\:cm^{-2}}$ clump (see \S2). Wright et al. (1992)
estimate a current core gas mass of about $17-38\:M_\odot$. This core
was also probed by its extinction in the 9.8~$\rm \mu m$ silicate
feature by Gezari et al. (1998), with the extinction peaking close to
source {\it I}. The polarization vectors of near to mid IR emission
suggest that a single source is responsible for much of the luminosity
from the core (Werner, Capps, \& Dinerstein 1983).

Wright et al. (1995) interpreted SiO (v=0; J=2-1) maser emission that
is centered about source {\it I} as indicating the presence of a
$r\sim 1000$~AU accretion disk (perhaps interacting with an outflow so
that motions are not precisely Keplerian).  This scale is similar to
that estimated from collapse of a core with $\beta=0.02$ (\S2). The
velocity of maser spots from different sides of the disk suggest a
central mass of about 20~$M_\odot$. There is apparent elongation along
the SW to NE axis because of the inclination of the disk with respect
to our line of sight. The disk alignment is perpendicular to the large
scale molecular outflow to the NW and SE (Chernin \& Wright 1996). The
apparent ``explosive'' appearance of the inner part of this outflow
(Allen \& Burton 1993) could be due to tidally-enhanced accretion and
accretion powered-outflow caused by the close passage of BN with
source {\it I} about 500 years ago (Tan 2004). It should be noted
that the outflow extends beyond the region that is usually considered
to be ``explosive'' (e.g. Henney et al. 2007), as would be expected
in this scenario where the protostar is $\sim 10^5\:{\rm yr}$ old.


To derive the properties of the protostar, we can consider the
bolometric luminosity coming from the KL core, $\sim 5\times
10^4\:L_\odot$ (Kaufman et al. 1998; Gezari et al. 1998), with an
uncertainty at about the factor of 2 level. Comparing to the
protostellar evolution models of McKee \& Tan (2003), one possible set
of parameters for the protostar is: $m_*=18\:M_\odot$,
$\dot{m}_*=3.6\times 10^{-4}:M_\odot\:{\rm yr^{-1}}$ and $S_*\sim 4\times
10^{47}\:{\rm s^{-1}}$ (Tan 2003).

The ionizing photons will interact primarily with the outflow gas as
it is magneto-centrifugally-launched up from the accretion disk,
creating an ``outflow-confined HII region'' (Tan \& McKee 2003). These
HII regions are unconfined in polar directions along the disk/outflow
rotation axis, and if the ionizing flux is strong enough, can become
unconfined in near-equatorial directions also. However, their emission
measure will always be strongly peaked around the protostar because of
the approximately $r^{-2}$ density profile in the outflow. Tan \& McKee (2003)
showed this model fits the radio spectrum of source {\it I} very well,
and naturally explains the observed elongation of the radio source
along the NW-SE axis (Reid et al. 2007), i.e.  parallel to that of the
larger scale outflow (Chernin \& Wright 1996). The position angle of
elongation aligns well with a particular Herbig-Haro object to the NW
(Taylor et al. 1986) requiring flow velocities $\sim 1000\:{\rm
km\:s^{-1}}$, which is about the escape speed and expected maximum
outflow velocity from a 20~$M_\odot$ protostar.

SiO (v=1 \& 2) masers have been observed surrounding the radio source
on scales of several tens of AU (Greenhill et al. 2004; Greenhill et
al., these proceedings). The densities and temperatures of the gas in
the outflow-confined HII region model are appropriate for the
excitation of these masers. However, the maser velocities are rather
low ($\sim 10-20\:{\rm km\:s^{-1}}$), although there are observational
biases against detecting relative velocities $\gtrsim 50\:{\rm
km\:s^{-1}}$ (Greenhill \& Matthews, priv. comm.).  A velocity
gradient is seen along the elongated direction of source {\it
I}. Greenhill et al. (2004) used this to argue that the disk is in
fact orientated along this axis, perpendicular to the previously
described disk model. However, there is little evidence for a large
outflow or outflow cavity in the direction expected for this
orientation (rather the opposite: a large amount of dense gas and
extinction; Gezari et al. 1998) and a new source would be needed for
the powerful NW-SE outflow. Possibilities to reconcile the observed
SiO (v=1 \& 2) maser motions with the outflow-confined HII region
model include: (1) the maser features may correspond to patterns of
temperature and density variation rather than actual gas motion; (2)
very particular orientations of protostar, gas and our line of sight
may be needed for maser amplification so that the full velocity field
is not sampled; (3) SiO emission may be limited inside the dust
destruction front (very few maser spots are seen with projected
distances $<10$~AU from the center of source {\it I}) and so the
observed spots may trace the kinematics of gas launched from
regions beyond the dust destruction front, where the escape speeds are
only about a few tens of $\rm km\:s^{-1}$.

\section{Conclusions}

It is remarkable that, given its astrophysical importance, massive
star formation remains so poorly understood. Theories that involve
basic differences in the accretion mechanism are actively
debated. After defining the terminology and physical properties
expected of cores forming together in a star-forming clump, we see
that the theoretical differences boil down to whether massive star
formation proceeds from massive gravitationally bound gas cores or
from competitive accretion to protostellar seeds that are already
well-formed before much of their gas is accumulated - i.e. in this
latter case the local core potential is mostly determined by the
stellar mass rather than the gas mass. We favor the theory of massive
star formation from massive gas cores, and suspect numerical models
will support this view once they include the physics of magnetic
fields (that can help support massive cores) and feedback from
protostellar outflows and radiation pressure (that inhibit Bondi-Hoyle
accretion and small scale fragmentation near massive protostars -
Krumholz 2006).

However, even in the context of models of formation from massive gas
cores, understanding what is going on in even the nearest massive
protostar and core remains challenging. The Orion KL region is crowded
with young stars, and close dynamical interactions definitely occur
between them, such as must have accelerated the BN object. I have
argued $\Theta^1C$ is responsible for this event, because it satisfies
all the required properties expected of the recoiling, eccentric,
hard, massive binary system that must be left behind. As proper motion
measurements improve, this issue should be resolved
definitively. Source {\it I} is likely to be an actively accreting
massive protostar, that was perturbed by BN's close passage. However,
even the orientation of the disk/outflow axis of this system is
debated from two orthogonal possibilities. Outflow-confined HII
regions are a prediction of massive star formation models that are
scaled-up versions of low-mass star formation models. The radio
spectrum and morphology of source {\it I} can be explained by this
type of model, but the kinematics of the excited SiO maser spots on
20-100~AU scales remain mysterious.



\acknowledgements 
JCT is supported by NSF CAREER grant AST-0645412.



\end{document}